\documentclass[Afour,sageh,times]{sagej}

\usepackage{moreverb,url,graphicx}
\usepackage{url}
\usepackage[colorlinks,citecolor=blue,urlcolor=blue]{hyperref}

\usepackage{fancyhdr}
\makeatletter
\def\@maketitle{%
\vspace*{-20pt}%
\null%
\begin{center}
\begin{sf}
\begin{minipage}[t]{\textwidth}
  \vskip 12.5pt%
    {\raggedright\titlesize\textbf{\@title} \par}%
    \vskip 1.5em%
    \vskip 12.5mm%
\end{minipage}
{\par\large%
      \vspace*{6mm}%
      \lineskip .5em%
      {\raggedright\textbf{\@author}
      \par}}
     \vskip 40pt%
    {\noindent\usebox\absbox\par}
    {\vspace{20pt}%
      {\noindent\normalsize\@keywords}\par}
\end{sf}
\end{center}
\vspace{22pt}
\par%
}
\makeatother

\def\volumeyear{}
\def\volumenumber{}
\def\issuenumber{}
\def\journalname{}

\begin{document}

\title{Reorienting Age-Friendly Frameworks for Rural Contexts: 
A Spatial Competence-Press Framework for Aging in Chinese Villages}

\author{Ziyuan Gao}

\affiliation{University College London}

\email{clairegao0930@gmail.com}

\begin{abstract}
While frameworks such as the WHO Age-Friendly Cities have advanced urban aging policy, rural contexts demand fundamentally different analytical approaches. The spatial dispersion, terrain variability, and agricultural labor dependencies that characterize rural aging experiences require moving beyond service-domain frameworks toward spatial stress assessment models. Current research on rural aging in China exhibits methodological gaps, systematically underrepresenting the spatial stressors that older adults face daily, including terrain barriers, infrastructure limitations, climate exposure, and agricultural labor burdens. Existing rural revitalization policies emphasize standardized interventions while inadequately addressing spatial heterogeneity and the spatially-differentiated needs of aging populations.
This study developed a GIS-based spatial stress analysis framework that applies Lawton and Nahemow's competence-press model to quantify aging-related stressors and classify rural villages by intervention needs. Using data from 27 villages in Mamuchi Township, Shandong Province, we established four spatial stress indicators: slope gradient index (SGI), solar radiation exposure index (SREI), walkability index (WI), and agricultural intensity index (AII). Analysis of variance and hierarchical clustering revealed significant variation in spatial pressures across villages and identified distinct typologies that require targeted intervention strategies. The framework produces both quantitative stress measurements for individual villages and a classification system that groups villages with similar stress patterns, providing planners and policymakers with practical tools for designing spatially-targeted age-friendly interventions in rural China and similar contexts.

\end{abstract}


\twocolumn[\maketitle]

%%%%%%%%%%%%%%%%%%%%%%%%%%%%%%%%%%%%%%%%%%%%%%%%%%%%%%%%%%%%%%%%%%%%%%%%%%%%%%%%%%%%

\section{Spatial Challenges in China's Rural Aging Context}

Rural areas in China are facing a significant demographic transformation. The latest data shows that the proportion of the older population aged 65 and above has exceeded 23\% in many rural villages, far surpassing the national average of 13.82\% \citep{NationalStats2023}. This aging trend exhibits significant regional variation, and institutional capacity deficits exacerbate challenges in care support for rural older populations. Rural older people generally face three interconnected spatial challenges: First, infrastructure is inadequate, and many villages lack age-friendly renovations; Second, physical capacity for agricultural activities and daily mobility, both essential components of rural life, declines with age. Thirdly, geographical isolation leads to a decrease in service accessibility \citep{HanEtAl2023}. These factors superimpose on each other, significantly magnifying the health risks and social vulnerabilities of the older population.

What is more worthy of attention is that the traditional family-based care model for older people is disintegrating. The outflow of young and middle-aged people and the subsequent impact of the one-child policy have made the phenomenon of ``empty-nest older people'' particularly pronounced in rural areas \citep{ZhanMontgomery2003}, where over 60.5\% of older adults are empty-nest older people \citep{ChangEtAl2016} and rural empty-nest older people are facing more serious health challenges than those in urban areas \citep{LiEtAl2022}. This demographic transformation compounds spatial vulnerabilities, as older adults with diminished family support networks must independently navigate challenging terrain, extreme weather exposure, and ongoing agricultural labor demands.

Meanwhile, the primary medical and health service system has substantial deficits in rural areas, where spatial isolation compounds service accessibility problems. Rural residents experience poorer health conditions and higher disease burden compared to urban residents but have lower healthcare services utilization \citep{ChenEtAl2023}, and rural older adults obtain less welfare compared to urban older adults due to inadequate social security benefits \citep{ZhangEtAl2023a, LiuEtAl2023a}. These institutional deficits, combined with spatial barriers, further exacerbate the marginalization of rural older people.

Although the aging challenge in rural areas of China is becoming increasingly severe, there is a significant disconnection between the existing policy framework and the actual needs of rural areas. Although national policies such as the China's Rural Revitalization Strategy (2025--2035) \citep{ChinaCouncil2022} have acknowledged the need to adapt rural environments for an aging society, the implementation remains uneven, inconsistent and often lacks spatial precision. The current mainstream framework guiding the older people service system, especially WHO Age-Friendly Cities Framework \citep{WHO2007}, was developed for urban contexts and cannot adequately assess the specific spatial challenges that rural older adults face in their daily lives. Therefore, there is an urgent need to develop measurement tools that can identify and quantify rural-specific aging stressors.

% ============================================================================
% LITERATURE REVIEW SECTION
% ============================================================================

\section{Current Research on Rural Aging in China: The Absence of Spatial Perspectives}

The discourse on aging has traditionally been dominated by urban-oriented paradigms, with the \citeauthor{WHO2007}'s (\citeyear{WHO2007}) Age-Friendly Cities Framework representing the most widely adopted global model. This framework's eight domains assume urban characteristics such as centralized public transit systems, walkable neighborhoods with sidewalks, and concentrated housing that enables efficient service delivery. However, rural communities often exhibit spatial dispersion, weak institutional capacity and livelihood dependence in specific contexts, characteristics that distinguish them from urban settings. Urban gerontology typically assumes access to formal care services, proximate healthcare facilities, and professional support networks, while rural contexts are characterized by informal care arrangements, geographic isolation from services, and agricultural livelihood dependencies \citep{Hash2015}. These distinctions make existing frameworks unsuitable for addressing the spatial and livelihood complexities of rural aging contexts. While \citeauthor{LawtonNahemow1973}'s (\citeyear{LawtonNahemow1973}) competence-press model has been extensively applied in urban settings, its application to rural contexts remains limited, failing to operationalize spatial ``press'' in ways that capture the unique rural stressors affecting older adults.

Research on rural aging in China has expanded recently but remains fragmented in scope and methodology. Many studies emphasize macro-level population trends \citep{ZengWang2014} or focus on psychosocial health of left-behind older population \citep{SilversteinEtAl2006}, with limited attention to village-scale environmental variability and spatial-specific impacts on aging experiences. While studies call for spatial-based approaches \citep{DouEtAl2025}, these have not been accompanied by systematic frameworks for measuring and categorizing spatial pressures that older adults face in daily rural contexts.

This fragmentation extends to village classification systems, which inadequately capture aging-relevant spatial characteristics. Building upon existing classification frameworks, effective age-friendly interventions require specialized typologies that group settlements by aging-relevant spatial characteristics to complement general development indicators. However, current classification systems fail to integrate aging-specific spatial stressors, as they employ overly broad categorizations based on GDP or administrative status \citep{ZhangPan2020} rather than aging-relevant factors such as slope gradients, walkability conditions, or agricultural labor intensity, lacking the granularity to support spatially differentiated aging interventions. For instance, the widely used classification of service-centered villages (SCVs) and agricultural-dependent villages (ADVs), where SCVs serve as administrative and infrastructure centers while ADVs are shaped by persistent agricultural labor demands and limited public services \citep{LiuEtAl2022}, was not designed with aging populations in mind and therefore does not address the spatial barriers that disproportionately affect older adults. Similarly, typological distinctions such as ``traditional agricultural villages'' or ``hollow villages'' offer limited insights into the complex interactions among terrain, infrastructure and livelihood practices. Recent studies have called for the adoption of environmentally sensitive and age-friendly approaches in rural China, emphasizing the need to align interventions with the functional and spatial characteristics of specific village types \citep{LiuEtAl2020a}. This study addresses this need by developing a spatial stress framework that assess and classifies villages based on aging-relevant characteristics.

Environmental gerontology research in China has not yet fully incorporated spatial indicators such as slope gradient, sunlight exposure, and path walkability to analyze their impact on rural aging. This deficiency is particularly evident in the aspect of agricultural labor burden. Although studies generally recognize the impact of farming on the lives of rural older people, there is a lack of systematic and spatialized assessment indicators, especially the measurement of agricultural intensity that can reflect the physical endurance of the older people. The seasonal nature of agricultural work, combined with manual labor demands, places significant pressure on older population, many of whom continue to engage in farming due to economic necessity \citep{LiuEtAl2023b}. Despite this, most analyses do not adequately address the temporally concentrated and physiologically intensive nature of such labor, nor do they account for village-level heterogeneity in cropping systems, mechanization, or labor substitution opportunities. This omission limits both the explanatory power of current models and their utility for guiding agricultural modernization or labor-relief interventions targeted at physically vulnerable aging populations.

The absence of integrated frameworks that can simultaneously measure multiple spatial stressors, classify villages by intervention needs, and connect to established aging theories represents a significant gap. This necessitates developing spatial frameworks that can quantify multiple spatial pressures, classify villages by aging-relevant characteristics, and guide targeted interventions.

% ============================================================================
% THEORETICAL FRAMEWORK
% ============================================================================

\section{A Competence--Press Framework for Rural Aging in China}

This study applies \citeauthor{LawtonNahemow1973}'s (\citeyear{LawtonNahemow1973}) Competence--Press Model to interpret spatial variation in rural aging experiences. The model emphasizes the dynamic fit between individual functional capacity and external environmental press. In rural areas of China, this fit is especially strained by terrain obstacles (measured through slope gradients), insufficient infrastructure (captured by walkability conditions), and labor-intensive livelihoods (quantified through agricultural intensity). Drawing from environmental gerontology principles, we operationalize environmental `press' through four spatially measurable indicators that capture the specific stressors faced by older adults at the village scale: terrain difficulty (slope gradients), climate exposure (solar exposure), mobility barriers (walkability), and agricultural labor demands (agricultural intensity).

Drawing on the competence-press theory, this study develops a dual-stage analytical framework. Stage 1 focuses on spatial stress quantification through four spatial stress indicators that represent the most critical stressors faced by older adults in rural village contexts. The Slope Gradient Index (SGI) measures terrain steepness that affects older adults' daily mobility and fall risks. The Solar Radiation Exposure Index (SREI) quantifies heat exposure that impacts older people's health and outdoor activity patterns. The Walkability Index (WI) evaluates path conditions that determine older adults' spatial accessibility within villages. The Agricultural Intensity Index (AII) measures farming workload demands that burden aging rural populations. These four indices are all generated through geospatial analysis via ArcGIS Pro, integrating terrain modeling, solar radiation simulation, path classification and agricultural system data.

Stage 2 adopts a systematic three-stage analytical approach to examine spatial stress patterns affecting aging populations in villages: First, correlation analysis and multicollinearity assessment were conducted to examine interdependence among four spatial stress indicators (SGI, SREI, WI, and AII). Second, Linear Discriminant Analysis was applied to evaluate whether existing administrative classifications effectively capture spatial stress variations, revealing suboptimal alignment between administrative categories and spatial stress patterns. Third, hierarchical clustering using Mahalanobis distance identified an optimal village classification system that demonstrated superior statistical performance compared to the binary administrative system. This comprehensive analytical framework of ``correlation assessment - discriminant validation - spatial clustering'' provides a robust scientific foundation for developing differentiated age-friendly interventions that address distinct spatial stress profiles rather than relying solely on administrative classifications.

The traditional urban-centered aging framework, including the WHO Age-Friendly Cities model and existing village classification systems like service-centered villages (SCVs) and agricultural-dependent villages (ADVs), is insufficient to address the spatial heterogeneity of rural residential areas, encompassing varied terrain conditions, dispersed infrastructure networks, diverse microclimates, and heterogeneous agricultural labor patterns. By integrating the competence-press model with GIS-based spatial analysis, this study develops a comprehensive framework that both quantifies aging-related spatial stressors and produces village typologies based on intervention needs, offering a more spatially-sensitive approach to rural aging assessment than existing administratively-defined classification systems.

% ============================================================================
% Case Selection Rationale
% ============================================================================
\section{Case Selection Rationale}
This study included all 27 villages under the jurisdiction of Mamuchi Township, Shandong Province (see Figure \ref{fig:gis}(a)). The selection provides an appropriate empirical setting for applying the Competence-Press Framework to rural aging contexts, offering the spatial heterogeneity and demographic characteristics necessary for comprehensive spatial stress analysis across diverse rural conditions.

\subsection{Economic Profile}
Mamuchi Township demonstrates a diversified economic structure with agriculture contributing 40\% of total output value. The red tourism industry, centered around the 4A-level Yimeng Red Film and Television Base, serves as a core growth pole \citep{ChinaRadio2018}, with the area designated as a 2022 Shandong Province Premium Cultural Tourism Town \citep{Tcmap2023}. Emerging business forms including polygonatum cultivation cooperatives and rural e-commerce have enhanced economic resilience \citep{Toutiao2022}. This ternary structure of "traditional agriculture + red cultural tourism + specialized cooperatives" represents typical industrial integration under China's rural revitalization strategy \citep{Shandong2022}.

\subsection{Demographic Aging Representativeness}
Mamuchi Township exhibits a significantly higher proportion of older people residents compared to the national rural average, with over 23\% of residents aged 65+ in 2022 compared to the national rural average of 13.82\% \citep{NationalStats2023}. This demographic profile reflects the accelerated aging characteristic of China's rural areas, often described as "getting old before getting rich" \citep{GaoEtAl2024} - where aging populations emerge before economic development can support adequate care infrastructure for older people.

\subsection{Topographic Heterogeneity}
Mamuchi Township exhibits diverse terrain features that provide an ideal setting for studying spatial pressures and mobility constraints among older adults. The township is predominantly mountainous with small riverside plains, consisting mainly of mountains and hills, with the highest peak, Wangjia'an North Mountain, reaching 618 meters above sea level. The selected villages represent distinct topographical conditions: Bama village features karst topography with over 42\% of its area exceeding 15° slopes; Zhangjiayuzi, located in the loess hilly area, has main slopes ranging from 8° to 25°; while Changshanzhuang represents a low-slope plain village with gradients consistently below 5° \citep{MinistryAffairs2015}. This topographic diversity enables comparative analysis of terrain-related mobility constraints among older adults.

\subsection{Policy Alignment}
Mamuchi Township has been designated pilot area under Shandong Province's Rural Revitalization Demonstration Project (2024-2027) \citep{ShandongGov2023} for age-friendly infrastructure adaptation. This ensures access to infrastructure upgrade data (e.g., gravel path coverage) and facilitates coordination with local administrative bodies for implementation support and data collection.

\section{Method}
Informed by Lawton and Nahemow's competence-press framework, this study developed a GIS-based methodology using a dual analytical approach. First, we quantified four spatial stress indicators (slope gradient, solar radiation exposure, walkability, and agricultural intensity) across 27 villages in Mamuchi Township using ArcGIS Pro and validated differences through ANOVA. Second, we applied hierarchical clustering to identify village typologies based on stress patterns. This dual approach enables both quantitative stress assessment for individual villages and classification systems that group villages with similar intervention needs, providing planners with practical tools for designing spatially-targeted age-friendly interventions.

% ============================================================================
% METHODS: SPATIAL STRESS INDICATORS
% ============================================================================
\begin{figure*}[!t]
    \centering
    \includegraphics[width=0.9\textwidth]{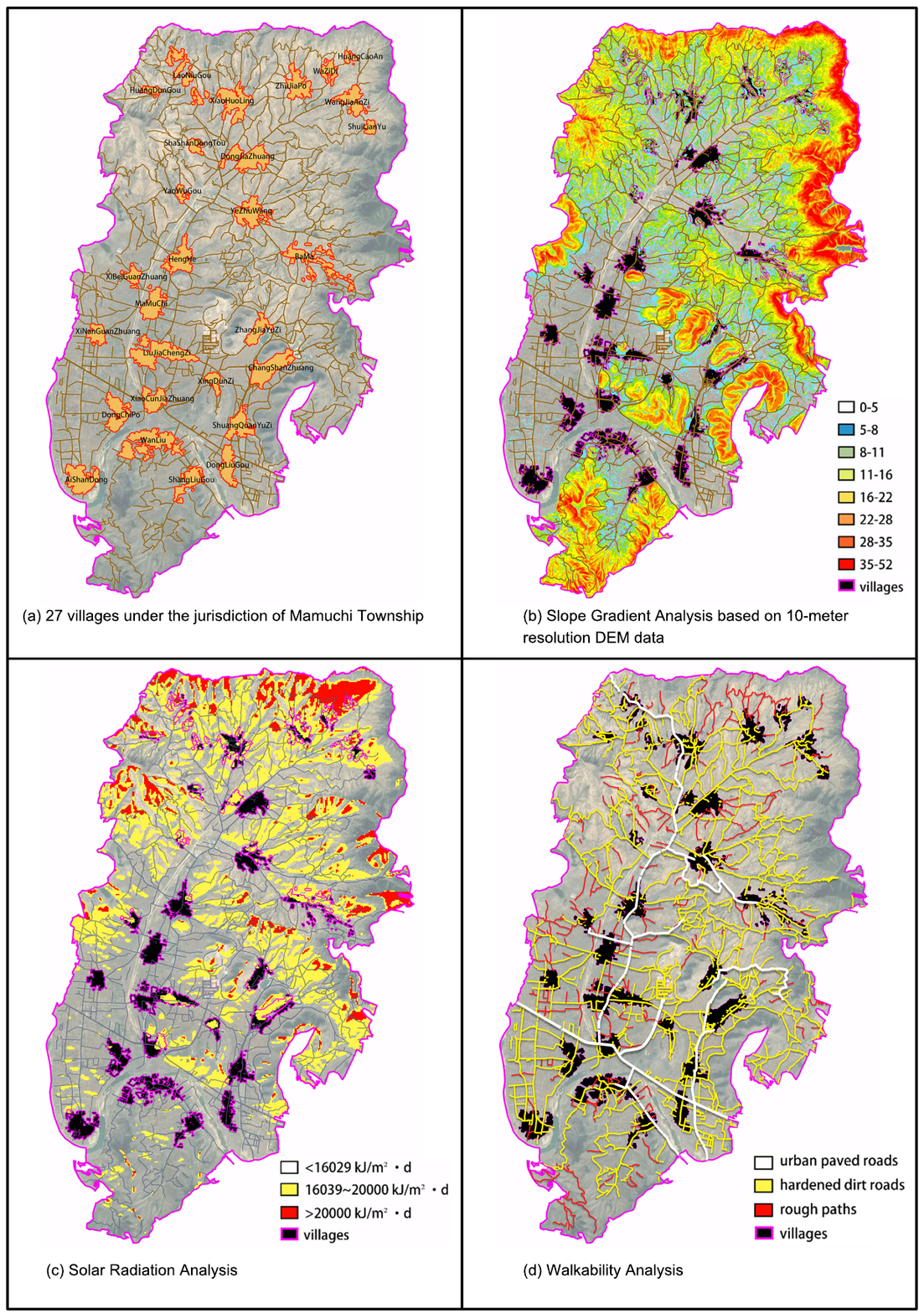} 
    \caption{Spatial analysis of four stress indicators across 27 villages in Mamuchi Township. (a) Geographic distribution of study villages showing administrative boundaries and settlement patterns. (b) Slope Gradient Index (SGI) derived from 10-meter resolution Digital Elevation Model (DEM), with red areas indicating steep terrain ($>4.8°$) posing mobility challenges for older adults. (c) Solar Radiation Exposure Index (SREI) showing areas exceeding the 20,000 kJ/m²·day threshold (red zones) where older residents face heat-related health risks. (d) Walkability Index (WI) illustrating road surface quality, with urban paved roads (white), hardened dirt roads (yellow), and rough paths (red) affecting older adults' spatial accessibility}
    \label{fig:gis}
\end{figure*}

\section{Stage 1: Spatial Stress Quantification}

\subsection{Slope Gradient Index (SGI)}

Given the age-related mobility limitations discussed above, we define the Slope Gradient Index (SGI) as the proportion of village area where slopes exceed the established $4.8°$ threshold, calculated as:

\begin{equation}
\text{SGI} = \frac{\text{area with slopes} > 4.8°}{\text{total village area}}
\end{equation}

This metric yields values from 0 (completely accessible terrain) to 1 (entirely inaccessible terrain). Using 10-meter resolution Digital Elevation Model (DEM) data (see Figure \ref{fig:gis}(b)), spatial analysis reveals dramatic variation across different topographical contexts: Zhangjiayuzi Village in the loess hilly area demonstrates an SGI of 0.84, indicating that 84\% of the village area presents mobility challenges, whereas Changshanzhuang Village in the plain area registers only 0.02, with merely 2\% of its area exceeding the critical slope threshold.

\subsection{Solar Radiation Exposure Index (SREI)}

Given the vulnerabilities discussed above, we define the Solar Radiation Exposure Index (SREI) to quantify solar radiation exposure risks in rural settings, particularly for older adults who are more susceptible to such exposure than younger populations. The index is quantified through solar radiation modeling using ArcGIS Pro's Area Solar Radiation tool (see Figure \ref{fig:gis}(c)). The index uses 20,000~kJ/m²·day as the critical threshold above which older adults face significant health risks, calculated as:

\begin{equation}
\text{SREI} = \frac{\text{area exceeding 20,000 kJ/m²·day}}{\text{total village area}}
\end{equation}

Values range from 0 (safe for older adult outdoor activities) to 1 (hazardous exposure levels).

The analysis of the village-level solar radiation Exposure Index (SREI) revealed significant spatial differentiation characteristics. The radiation values in villages such as Zhangjiayuzi Village (0.93), Huangcao'an Village (0.91), and Zhujiapo Village (0.72) far exceeded the threshold, forming continuous high-exposure areas, exposing the local older people to the health risks of combined exposure to thermal radiation and ultraviolet rays. In contrast, in low-value areas such as Dongliugou (0.00), Mamuchi (0.00), and Aishandong (0.02), the radiation intensity was significantly weakened by the micro-terrain shielding effect. This spatial heterogeneity indicates that there are regional differences in the risk of solar radiation exposure.

\subsection{Walkability Index (WI)}

In the research on rural aging, the quality of walking infrastructure is a key factor affecting the activity ability of the older people. The Walkability Index (WI) calculates the age-friendly level of the village road network by multiplying the percentage of each surface type by the specified material index value. This index adopts a three-level weighted calculation system: asphalt/concrete urban paved roads are assigned a value of 1.0, hardened dirt roads are 0.6, and unhardened dirt roads or rough paths are 0.3 (see Figure \ref{fig:gis}(d)). This classification standard reflects the negative correlation between road surface stability and the risk of falls among the older people \citep{WebberEtAl2010}. Mathematically:

\begin{equation}
\begin{split}
\text{WI} = & (\text{Urban Paved \%} \times 1.0) \\
            & + (\text{Hardened Dirt \%} \times 0.6) \\
            & + (\text{Rough Trail \%} \times 0.3)
\end{split}
\end{equation}

Villages such as Henghe (0.80), Shuangquanyuzi (0.79), and Cuijiazhuang (0.75) offer the most walkable environments, characterized by higher shares of paved or moderately surfaced routes. These conditions support safer walking and facilitate continued social and service engagement for older adults. In contrast, Zhujiapo (0.48), Wazidi (0.48), and Aishandong (0.50) fall at the lower end of the spectrum, with infrastructure dominated by unpaved or rough paths that may hinder mobility and discourage outdoor activity among aging residents.

In line with ecological models of aging, walkability acts as a key modifiable environmental press---shaping how older residents interact with their daily environment \citep{LawtonNahemow1973}. The WI thus offers a practical and spatialized tool for identifying mobility barriers and informing targeted infrastructure improvements.

\subsection{Agricultural Intensity Index (AII)}

Given the vulnerabilities discussed above, we define the Agricultural Intensity Index (AII) to quantify agricultural labor pressures in rural settings, particularly given that older adults are more vulnerable to intensive physical demands than younger populations. The index combines three weighted components: crop-specific labor intensity (1.0 for manually harvested cash crops, 0.6 for orchard crops, 0.3 for mechanized cultivation), temporal concentration (reflecting harvest period overlap), and labor force distribution (village-level agricultural dependency), calculated as a multiplicative composite ranging from 0 to 1.

Data for the Agricultural Intensity Index were collected from Yinan County Government reports (2018--2023) and verified media sources \citep{Shandong2022, ChinaRadio2018}. Analysis revealed significant spatial variation: villages specializing in labor-intensive dual-crop systems (e.g., Zhangjiayuzi's ginger-rosemary, AII$=$0.6) exhibit 3.3$\times$ greater seasonal pressure than single-crop grain communities (e.g., Wazidi's wheat-corn, AII$=$0.18). Crucially, analysis revealed that 68\% of villages have moderate-to-high agricultural intensity ($\geq$0.36), occurring in a township where over 23\% of residents are aged 65+. This creates challenges where aging populations must perform intensive agricultural labor.

% ============================================================================
% RESULTS: CORRELATION AND DISCRIMINANT ANALYSIS
% ============================================================================
\section{Stage 2: Spatial Stress Analysis and Village Typology Development}

This study developed a village typology through a three-stage analytical process: correlation assessment, discriminant analysis, and clustering validation. The approach ensures that the resulting classification system optimally captures spatial stress patterns affecting aging populations while maintaining statistical robustness.

\subsection{Stage 2.1: Correlation Analysis and Multicollinearity Assessment}

\begin{figure}[htbp]
    \centering
    \includegraphics[width=0.8\columnwidth]{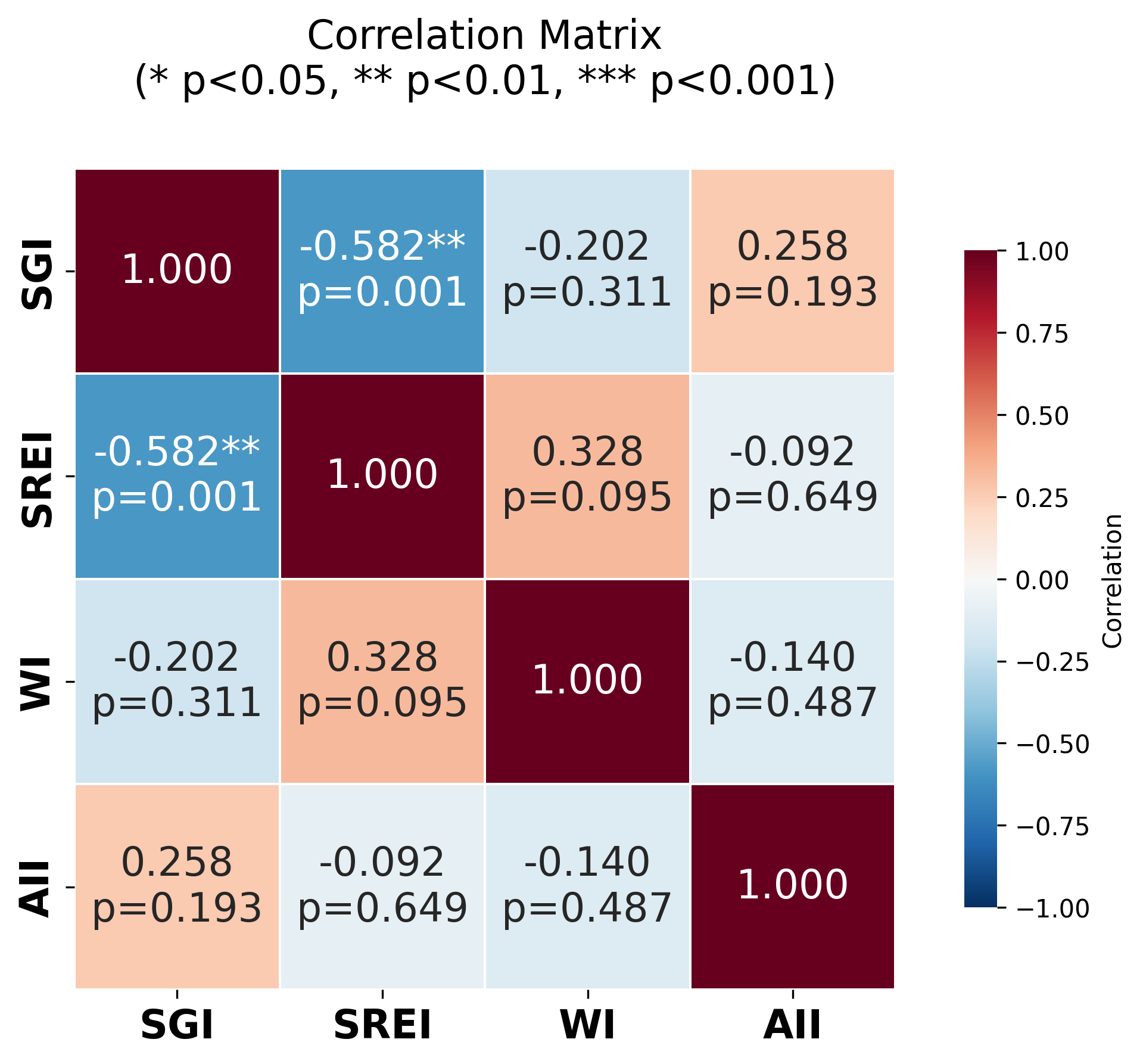}
    \caption{Correlation Matrix of Four Spatial Stress Indicators}
    \label{fig:correlation_heatmap}
\end{figure}

Correlation analysis examined interdependence among the four spatial stress indicators using Pearson correlation coefficients and Variance Inflation Factor (VIF) analysis (Figure \ref{fig:correlation_heatmap}). The correlation matrix revealed moderate negative correlation between SGI and SREI ($-0.582$), positive correlation between SREI and WI (0.328), and weak positive correlation between SGI and AII (0.258).

Among 6 indicator pairs, only 1 pair (SGI-SREI) showed significant moderate negative correlation ($r=-0.5816$, $p=0.0015$), revealing meaningful inverse relationship between terrain gradient and solar exposure. The remaining pairs showed weak to moderate correlations without statistical significance. Therefore, the four spatial stress indicators are not completely independent.

\subsection{Stage 2.2: Discriminant Analysis for Administrative Classification Evaluation}

This function creates optimal linear combinations to distinguish between ADV and SCV villages, with Walkability Index showing the strongest positive contribution (15.0831) and Agricultural Intensity Index showing strong negative contribution ($-8.1405$).

To evaluate whether the traditional administrative classifications align with spatial stress patterns, we applied the discriminant function to test how well spatial stress indicators can predict administrative categories (ADV vs SCV). The discriminant scores (LD1) represent each village's position along the discriminant dimension, where negative values indicate ADV-like characteristics and positive values indicate SCV-like characteristics based on spatial stress patterns. According to the discriminant analysis results (Table \ref{T1}), the moderate 74.07\% accuracy (20 correct out of 27 villages) indicates that administrative classification does not optimally capture spatial stress variations. The overlapping LD1 ranges between groups (ADV max: 0.941 $>$ SCV min: $-0.338$) explain the classification errors, suggesting that some villages are misaligned with their administrative categories based on spatial stress patterns.

\begin{table*}[t]
\small\sf\centering
\caption{Administrative Classification Evaluation via Discriminant Analysis\label{T1}}
\begin{tabular}{lcccccc}
\toprule
Administrative Category & Count & LD1 Mean & LD1 SD & LD1 Range & Correct Predictions & Accuracy\\
\midrule
Agricultural-Dependent Village (ADV) & 15 & $-$0.759 & 1.065 & [$-$2.402, 0.941] & 12/15 & 80.0\%\\
Service-Centered Villages (SCV) & 12 & 0.949 & 0.911 & [$-$0.338, 2.177] & 8/12 & 66.7\%\\
Total & 27 &  &  &  & 20/27 & 74.07\%\\
\bottomrule
\end{tabular}\\[2pt]
{\flushleft\small Notes: Classification rule: LD1 $<$ 0 $\rightarrow$ ADV, LD1 $>$ 0 $\rightarrow$ SCV.\par}
\end{table*}

\begin{table*}[t]
\small\sf\centering
\caption{Statistical Comparison of Spatial Stress Indicators by Administrative Clustering and Spatial-based Clustering\label{T2}}

% 上半部分 - 列宽较宽
\setlength{\tabcolsep}{17pt}
\begin{tabular}{@{}l*{4}{c}ccc@{}}
\toprule
\multicolumn{8}{c}{\textbf{Type of villages (Administrative Clustering)}}\\
\midrule
 & \multicolumn{2}{c}{ADV (N=15)} & \multicolumn{2}{c}{SCV (N=12)} & \makebox[2.5cm][c]{95\%CI} & \makebox[1.3cm][c]{F-Score} & \makebox[1.3cm][c]{P-value}\\
\cmidrule(lr){2-3} \cmidrule(lr){4-5}
 & Mean & SD & Mean & SD & & &\\
\midrule
SGI & 0.450 & 0.311 & 0.242 & 0.165 & [0.016, 0.400] & 4.385* & 0.047\\
SREI & 0.578 & 0.229 & 0.677 & 0.219 & [$-$0.277, 0.079] & 1.278 & 0.269\\
WI & 0.595 & 0.067 & 0.673 & 0.090 & [$-$0.142, $-$0.014] & 6.660* & 0.016\\
AII & 0.394 & 0.151 & 0.248 & 0.117 & [0.040, 0.252] & 7.645* & 0.011\\
\midrule
\end{tabular}

\vspace{1pt}

% 下半部分 - 列宽较窄
\setlength{\tabcolsep}{11.5pt}
\begin{tabular}{@{}l*{8}{c}cc@{}}
\multicolumn{11}{c}{\textbf{Type of villages (Spatial-based Clustering)}}\\
\midrule
 & \multicolumn{2}{c}{\makebox[1.2cm][c]{Cluster 1 (N=4)}} & \multicolumn{2}{c}{\makebox[1.2cm][c]{Cluster 2 (N=7)}} & \multicolumn{2}{c}{\makebox[1.2cm][c]{Cluster 3 (N=3)}} & \multicolumn{2}{c}{\makebox[1.3cm][c]{Cluster 4 (N=13)}} & \makebox[1.2cm][c]{F-Score} & \makebox[1.2cm][c]{P-value}\\
\cmidrule(lr){2-3} \cmidrule(lr){4-5} \cmidrule(lr){6-7} \cmidrule(lr){8-9}
 & Mean & SD & Mean & SD & Mean & SD & Mean & SD & &\\
\midrule
SGI & 0.512 & 0.181 & 0.594 & 0.237 & 0.013 & 0.012 & 0.262 & 0.212 & 7.58** & 0.0011\\
SREI & 0.370 & 0.162 & 0.458 & 0.180 & 0.873 & 0.033 & 0.730 & 0.149 & 11.16*** & 0.0001\\
WI & 0.575 & 0.075 & 0.569 & 0.050 & 0.557 & 0.051 & 0.695 & 0.061 & 10.33*** & 0.0002\\
AII & 0.158 & 0.045 & 0.463 & 0.108 & 0.324 & 0.062 & 0.292 & 0.117 & 7.97*** & 0.0008\\
\bottomrule
\end{tabular}
{\flushleft\small Note: *** $p < 0.001$, ** $p < 0.01$, * $p < 0.05$. SGI = Slope Gradient Index, SREI = Solar Radiation Exposure Index, WI = Walkability Index, AII = Agricultural Intensity Index.\par}
\end{table*}

% ============================================================================
% RESULTS: CLUSTERING ANALYSIS
% ============================================================================

\subsection{Stage 2.3: Spatial-based Clustering Implementation}

To establish a more effective village classification system based on spatial stress patterns, hierarchical clustering with average linkage was applied using the Mahalanobis's distance on four key spatial indicators (SGI, SREI, WI, and AII) after data transformation. This approach better accounts for the multivariate nature of spatial stress factors than the binary administrative classification.

We evaluated multiple clustering solutions ($k=2$ to $k=16$) using several validity indices. While the highest silhouette coefficient was observed at $k=4$, the Calinski-Harabasz (CH) index suggested stronger separation at $k=2$ (17.961) and $k=4$ (10.570). After careful evaluation of both statistical indices and practical interpretability, we selected the $k=4$ solution, which identified four distinct village types with significant differences across all four spatial stress indicators (all $p<0.01$). This four-cluster solution provides a more nuanced classification than the binary administrative approach, capturing meaningful variations in the spatial stress profiles that directly impact aging populations.

\subsection{Comparative Classification Validation}

To evaluate the effectiveness of the spatial-based classification against the existing administrative system (Table \ref{T2}), we conducted statistical comparisons of spatial stress indicators across different classification schemes. These analyses compare how the four spatial stress indicators differ between groups under each classification method, with F-scores and $p$-values indicating the statistical significance of differences between groups.

The spatial-based classification demonstrates markedly superior statistical performance. While administrative classification achieved significance for only 3 out of 4 indicators with moderate F-scores (4.385--7.645), the spatial-based approach achieved significance across all 4 indicators with substantially higher F-scores (SGI: 7.58, SREI: 11.16, WI: 10.33, AII: 7.97) and enhanced significance levels ($p < 0.001$ for SREI, WI, and AII; $p < 0.01$ for SGI). This indicates that the four-cluster spatial-based classification captures more distinct and statistically meaningful village groupings based on spatial stress patterns.

\subsection{Cross-Classification Analysis Between Administrative and Spatial-based Classifications}

We performed cross-classification analysis to examine how administrative categories align with spatial-based clusters. The analysis reveals important differences between the two administrative types. The distribution of villages shows that administrative designation doesn't consistently align with spatial characteristics. ADV villages are spread across all four spatial clusters, with concentrations in Clusters 2 and 4 (6 villages each), indicating significant spatial diversity within this administrative category. SCV villages show a notable concentration in Cluster 4 (7 villages), suggesting that service villages tend toward favorable spatial conditions but still exhibit some variation. This indicates administrative classification doesn't fully capture the spatial characteristics relevant to aging populations, highlighting the importance of spatial-based approaches for aging-friendly rural planning.

\begin{sidewaystable*}[p]
\small\sf\centering
\caption{Spatial-based Classification Cluster Characteristics and Interpretation\label{T3}}
\begin{tabular}{@{}p{4cm}p{3cm}p{5.5cm}p{4cm}p{4.5cm}@{}}
\toprule
\textbf{Cluster} & \textbf{Villages} & \textbf{Key Characteristics} & \textbf{Spatial Stress Profile} & \textbf{Aging Population Implications}\\
\midrule
\textbf{Cluster 1:} Steep Terrain + High Agricultural Load (N=4) 
& 
\begin{tabular}[t]{@{}l@{}}
Zhangjiayuzi\\
Yaowugou\\
Xingdunzi\\
Wazidi
\end{tabular}
& 
\begin{tabular}[t]{@{}l@{}}
High terrain gradient (0.662)\\[3pt]
Low solar exposure (0.436)\\[3pt]
Moderate walkability (0.603)\\[3pt]
Highest agricultural intensity (0.600)
\end{tabular}
& Steep topography constrains accessibility and infrastructure development 
& Mobility challenges due to topographic barriers and intensive farming demands\\[10pt]
\midrule
\textbf{Cluster 2:} Moderate Terrain + Low Solar Exposure + Moderate Agricultural Load (N=7) 
& 
\begin{tabular}[t]{@{}l@{}}
Bama Village\\
Wangjia'anzi\\
Shashandongtou\\
Huangdungou\\
Zhujiapo\\
Huangcao'an\\
Shuilianyu
\end{tabular}
& 
\begin{tabular}[t]{@{}l@{}}
Moderate terrain gradient (0.509)\\[3pt]
Lowest solar exposure (0.420)\\[3pt]
Lower walkability (0.553)\\[3pt]
Low agricultural intensity (0.244)
\end{tabular}
& Moderate topographic constraints with limited solar exposure 
& Mixed challenges requiring balanced interventions for aging residents\\[10pt]
\midrule
\textbf{Cluster 3:} Flat Terrain + High Solar Exposure + Low Agricultural Load (N=3) 
& 
\begin{tabular}[t]{@{}l@{}}
Xinan'guanzhuang\\
Dongchipuo\\
Aishandong
\end{tabular}
& 
\begin{tabular}[t]{@{}l@{}}
Very flat terrain (0.013)\\[3pt]
Highest solar exposure (0.873)\\[3pt]
Moderate walkability (0.557)\\[3pt]
Moderate agricultural intensity (0.324)
\end{tabular}
& Highly accessible terrain but significant heat/sun exposure challenges 
& Heat stress management needed while leveraging flat terrain advantages\\[10pt]
\midrule
\textbf{Cluster 4:} Gentle Terrain + Good Solar + Best Walkability (N=13) 
& 
\begin{tabular}[t]{@{}l@{}}
Changshanzhuang\\
Shuangquanyuzi\\
Dongliugou\\
Shangliugou\\
Cuijiazhuang\\
Dongjiazhuang\\
Henghe\\
Yezhuwang\\
Wanliuzhuang\\
Liujiachengzi\\
Mamuchi\\
Xiaohuoling\\
Xibeiguanzhuang
\end{tabular}
& 
\begin{tabular}[t]{@{}l@{}}
Low terrain gradient (0.262)\\[3pt]
High solar exposure (0.730)\\[3pt]
Best walkability (0.695)\\[3pt]
Low agricultural intensity (0.292)
\end{tabular}
& Most balanced spatial stress profile with gentle topography and good infrastructure 
& Optimal conditions for aging populations requiring minimal age-friendly interventions\\
\bottomrule
\end{tabular}
\end{sidewaystable*}
% ============================================================================
% DISCUSSION
% ============================================================================

\section{Implications and Future Directions}

This study reorients aging-friendly frameworks by developing a GIS-based spatial stress approach that offers significant advancement beyond the WHO Age-Friendly Cities model. The WHO's eight domains were developed primarily for urban contexts and may not fully account for topographical challenges and distributed service models. Our three-step method, consisting of stress identification, village classification, and targeted intervention, diverges by directly quantifying environmental pressures in dispersed rural settings. The four indicators (SGI, SREI, WI, AII) demonstrated significant spatial variation across 27 villages. Spatial-based clustering achieved superior statistical performance over administrative classification, with F-scores of 7.58--11.16 versus 4.385--7.645 and $p < 0.001$ for three indicators (SREI, WI, and AII). This approach expands on WHO's service-focused domains by capturing terrain adaptability challenges that largely remain unaddressed in current age-friendly planning.

The approach reveals structural deficiencies in current aging-related policies. China's Rural Revitalization Strategy (2025--2035) exemplifies this issue by predominantly adopting standardized interventions like building older people's activity centers and road paving while overlooking regional stress heterogeneity \citep{ChinaCouncil2022}. Our typology enables targeted interventions tailored to different village types. As shown in Table \ref{T3}, high-stress villages (Cluster 2) require comprehensive terrain modification and agricultural labor relief that could be piloted through special subsidies for older farmers. These subsidies should promote light agricultural machinery, including mini seeders and harvesting aids, and seasonal labor sharing mechanisms, particularly in villages with high agricultural intensity. Villages with favorable walkability conditions (Cluster 4) could serve as models for progressive age-friendly development. Isolated villages in steep terrain (Cluster 1) need focused infrastructure connections, while high solar exposure settlements (Cluster 3) require heat mitigation strategies alongside their topographical advantages.

Several limitations constrain the approach's generalizability and implementation. Index weights require calibration across different ecological contexts, as mountainous regions may exhibit different stress-adaptation relationships than plains. The Agricultural Intensity Index presents implementation challenges, demanding longitudinal data on seasonal labor patterns currently unavailable in administrative systems. Solar radiation exposure effects require epidemiological validation across diverse climate zones. Cross-departmental coordination mechanisms for implementing multi-indicator interventions remain underdeveloped.

Future research should prioritize cross-regional validation studies to establish context-specific weighting systems. Development of streamlined data collection protocols for AII measurement is essential. Establishing a longitudinal and multilevel database of agricultural labor intensity will support broader implementation. The approach provides methodological foundation for spatially-explicit rural aging assessment that can be systematically adapted across diverse rural contexts globally, enabling more precise age-friendly interventions in rapidly aging rural societies.

%%%%%%%%%%%%%%%%%%%%%%%%%%%%%%%%%%%%%%%%%%%%%%%%%%%%%%%%%%%%%%%%%%%%%%%%%%%%%%%%%%%%%%%
\section*{Acknowledgements}
I would like to express my appreciation to Prof. Deshun Zhang for sharing case information and supporting data collection.

\section*{Declaration of Conflicting Interests}
The author(s) declared no potential conflicts of interest with respect to the research, authorship, and/or publication of this article.

\section*{Funding}
The author(s) received no financial support for the research, authorship, and/or publication of this article.


\begin{thebibliography}{99}

\bibitem[ADA(2010)]{ADA2010}
ADA (2010) 2010 ADA Standards for Accessible Design. U.S. Department of Justice. Available at: \url{https://www.ada.gov} (accessed 20 April 2025).

% \bibitem[Beard JR, Officer A, de Carvalho IA, Sadana R, Pot AM, Michel JP et al.(2016)]{BeardEtAl2016}
% Beard JR, Officer A, de Carvalho IA, Sadana R, Pot AM, Michel JP et al. (2016) The World report on ageing and health: a policy framework for healthy ageing. The Lancet 387(10033): 2145–2154.

% \bibitem[Cerin E, Nathan A, van Cauwenberg J, Barnett DW and Barnett A(2017)]{CerinEtAl2017}
% Cerin E, Nathan A, van Cauwenberg J, Barnett DW and Barnett A (2017) The neighborhood physical environment and active travel in older adults: a systematic review and meta-analysis. International Journal of Behavioral Nutrition and Physical Activity 14(1): 15.

% \bibitem[Chang X, Ma L, Cui X, Tao T and Zhao S(2024)]{ChangEtAl2024}
% Chang X, Ma L, Cui X, Tao T and Zhao S (2024) Spatial equity evaluation of rural eldercare service resources based on accessibility: A case study of Huanxian County of Gansu Province, China. Chinese Geographical Science 34(5): 869–885.

\bibitem[Chang Y, Guo X, Guo L, Li Z, Yang H, Yu S, et al.(2016)]{ChangEtAl2016}
Chang Y, Guo X, Guo L, Li Z, Yang H, Yu S, et al. (2016) Comprehensive comparison between empty nest and non-empty nest elderly: A cross-sectional study among rural populations in northeast China. International Journal of Environmental Research and Public Health 13(9): 857.

\bibitem[Chen L, Wang Y and Li X(2023)]{ChenEtAl2023}
Chen L, Wang Y and Li X (2023) Impact of pension income on healthcare utilization of older adults in rural China. International Journal for Equity in Health 22: 162.

\bibitem[China Meteorological Administration(2024)]{ChinaMet2024}
China Meteorological Administration (2024) National Meteorological Information Center – China Meteorological Data Service Center. Available at: \url{https://data.cma.cn/} (accessed 28 April 2025).

\bibitem[China Radio International(2018)]{ChinaRadio2018}
China Radio International (2018) Yinan County, Linyi City, Shandong Province: Mamuchi Township's ``Three-Step'' strategy builds a harmonious Bama Village. Available at: \url{https://eco.cri.cn/chinanews/20180807/d44ef690-bf6b-b0ca-ced4-a793abe86fc1.html} (accessed 20 April 2025).

\bibitem[China State Council(2022)]{ChinaCouncil2022}
China State Council (2022) China's Rural Revitalization Strategy (2025–2035). Beijing: The State Council.

% \bibitem[Cohen J(1988)]{Cohen1988}
% Cohen J (1988) Statistical power analysis for the behavioral sciences (2nd ed.). Hillsdale, NJ: Lawrence Erlbaum Associates.

\bibitem[Dou H, Wang C, Cheng G, Ma L, Li Z, Chen S and Liu Y(2025)]{DouEtAl2025}
Dou H, Wang C, Cheng G, Ma L, Li Z, Chen S and Liu Y (2025) The vision of younger-seniors-based elderly care in rural China: based on population aging predictions from 2020 to 2050. Humanities and Social Sciences Communications 12: 669.

% \bibitem[Fang J and Bloom G(2010)]{FangBloom2010}
% Fang J and Bloom G (2010) China's rural health system and environment-related health risks. Journal of Contemporary China 19(63): 23–35.

\bibitem[Gao M, Jiang F, Wang J et al.(2024)]{GaoEtAl2024}
Gao M, Jiang F, Wang J et al. (2024) Population ageing and income inequality in rural China: an 18-year analysis. Humanities and Social Sciences Communications 11: 1605.

\bibitem[Han S, Li Y, Wu Q et al.(2023)]{HanEtAl2023}
Han S, Li Y, Wu Q et al. (2023) Challenges and responses of left-behind elderly and children in rural China. China CDC Weekly 5(22): 493–497.

\bibitem[Hash KM(2015)]{Hash2015}
Hash KM (2015) Aging in rural places: Policies, programs, and professional practice. New York, NY: Springer Publishing Company.

\bibitem[Hondula DM, Davis RE, Leisten MJ, Saha MV, Veazey LM and Wegner CR(2012)]{HondulaEtAl2012}
Hondula DM, Davis RE, Leisten MJ, Saha MV, Veazey LM and Wegner CR (2012) Fine-scale spatial variability of heat-related mortality in Philadelphia County, USA, from 1983-2008: a case-series analysis. Environmental Health 11(1): 16.

\bibitem[Kenney WL and Munce TA(2003)]{KenneyMunce2003}
Kenney WL and Munce TA (2003) Invited review: aging and human temperature regulation. Journal of Applied Physiology 95(6): 2598-2603.

\bibitem[Kenny GP, Yardley J, Brown C, Sigal RJ and Jay O(2010)]{KennyEtAl2010}
Kenny GP, Yardley J, Brown C, Sigal RJ and Jay O (2010) Heat stress in older individuals and patients with common chronic diseases. Canadian Medical Association Journal 182(10): 1053–1060.

% \bibitem[Langya News Network(2022)]{Langya2022}
% Langya News Network (2022) Linyi City Yinan County Mamuchi Township: ``Red-Green-Gold'' paints a colorful rural landscape. Available at: \url{https://baijiahao.baidu.com/s?id=1740739028630389748} (accessed 20 April 2025).

\bibitem[Lawton MP and Nahemow L(1973)]{LawtonNahemow1973}
Lawton MP and Nahemow L (1973) Ecology and the aging process. In: Eisdorfer C and Lawton MP (eds) The psychology of adult development and aging. Washington, DC: American Psychological Association, pp.619–674.

\bibitem[Li M, Zhang Y, Zhang Z, Zhang Y, Zhou L and Chen K(2022)]{LiEtAl2022}
Li M, Zhang Y, Zhang Z, Zhang Y, Zhou L and Chen K (2022) Seeking medical services among rural empty-nest elderly in China: a qualitative study. BMC Geriatrics 22(1): 207.

% \bibitem[Liao L and Gao X(2018)]{LiaoGao2018}
% Liao L and Gao X (2018) Spatial differentiation and governance typologies of rural functional areas in China. Progress in Geography 37(5): 594–603.

\bibitem[Liu C, Zhang X and Wang J(2020)]{LiuEtAl2020a}
Liu C, Zhang X and Wang J (2020) Prevalence and determinants of basic functional difficulties among older adults in rural China. BMC Geriatrics 20: 359.

\bibitem[Liu H, Wang J and Li M(2023)]{LiuEtAl2023a}
Liu H, Wang J and Li M (2023) Impact of the new rural social pension insurance on the health of the rural older adult population. BMC Public Health 23: 2156.

\bibitem[Liu J, Fang Y, Wang G, Liu B and Wang R(2023)]{LiuEtAl2023b}
Liu J, Fang Y, Wang G, Liu B and Wang R (2023) The aging of farmers and its challenges for labor-intensive agriculture in China: A perspective on farmland transfer plans for farmers' retirement. Journal of Rural Studies 100: 103013.

\bibitem[Liu J, Fang Y, Wang R and Zou C(2022)]{LiuEtAl2022}
Liu J, Fang Y, Wang R and Zou C (2022) Rural typology dynamics and drivers in peripheral areas: A case of Northeast China. Land Use Policy 120: 106260.

\bibitem[Liu Y, Li J and Yang Y(2021)]{LiuEtAl2021a}
Liu Y, Li J and Yang Y (2021) Strategic adjustment of land use policy under the economic transformation: A case study of Yangtze River Delta in China. Land Use Policy 72: 26-37.

\bibitem[Liu Y, Wu F and Li Z(2021)]{LiuEtAl2021b}
Liu Y, Wu F and Li Z (2021) Aging in place in rural China: Mobility barriers and health inequalities. Journal of Rural Studies 85: 182–191.

% \bibitem[Long H and Liu Y(2016)]{LongLiu2016}
% Long H and Liu Y (2016) Rural restructuring in China: Theory, practice and policy implications. Land Use Policy 55: 1–13.

% \bibitem[Long H, Liu Y, Wu X and Dong G(2016)]{LongEtAl2016}
% Long H, Liu Y, Wu X and Dong G (2016) Spatial differentiation of rural development in China: Patterns, drivers and policy implications. Applied Geography 66: 1–12.

\bibitem[Lucas R, McMichael AJ, Smith WT and Armstrong BK(2006)]{LucasEtAl2006}
Lucas R, McMichael AJ, Smith WT and Armstrong BK (2006) Solar ultraviolet radiation: Global burden of disease from solar UV radiation. Environmental Burden of Disease Series, No.13. Geneva: World Health Organization.

\bibitem[Ministry of Civil Affairs of the People's Republic of China(2015)]{MinistryAffairs2015}
Ministry of Civil Affairs of the People's Republic of China (2015) Administrative divisions compendium of the People's Republic of China: Shandong Province volume (L. Li, Gen. Ed.; X. Chen, Vol. Ed.). Beijing: China Social Publishing House.

% \bibitem[Ministry of Housing and Urban-Rural Development of the PRC(2012)]{MinistryHousing2012}
% Ministry of Housing and Urban-Rural Development of the PRC (2012) Code for design of accessibility of urban roads and buildings (GB 50763-2012). Beijing: China Architecture \& Building Press.

\bibitem[Musselwhite C(2017)]{Musselwhite2017}
Musselwhite C (2017) Creating age-friendly transport systems. In: Harper S (ed.) International Handbook on Ageing and Public Policy. Cheltenham: Edward Elgar Publishing, pp.273–287.

\bibitem[National Bureau of Statistics of China(2023)]{NationalStats2023}
National Bureau of Statistics of China (2023) China Statistical Yearbook 2022. Beijing: China Statistics Press.

% \bibitem[Qian J and Ramesh M(2023)]{QianRamesh2023}
% Qian J and Ramesh M (2023) Strengthening primary health care in China: governance and policy challenges. Health Economics, Policy and Law 19(1): 57-72.

\bibitem[Shandong News Network(2022)]{Shandong2022}
Shandong News Network (2022) Yinan County Mamuchi Township: ``Partnership'' revitalization, growing toward the sun. Available at: \url{https://www.sdnews.com.cn/sd/sdgd/202212/t20221214\_4149067.htm} (accessed 20 April 2025).

\bibitem[Shandong Provincial People's Government(2023)]{ShandongGov2023}
Shandong Provincial People's Government (2023) Shandong Province Rural Revitalization Demonstration Project (2024-2027). Jinan: Shandong Provincial People's Government.

\bibitem[Silverstein M, Cong Z and Li S(2006)]{SilversteinEtAl2006}
Silverstein M, Cong Z and Li S (2006) Intergenerational transfers and living arrangements of older people in rural China: Consequences for psychological well-being. The Journals of Gerontology Series B: Psychological Sciences and Social Sciences 61(5): S256–S266.

% \bibitem[State Council of China(2025)]{StateCouncil2025}
% State Council of China (2025) National Rural Revitalization Strategy (2024–2027). Beijing: The State Council. Available at: \url{https://www.gov.cn/zhengce/202501/content\_7000493.htm} (accessed 10 April 2025).

\bibitem[Tcmap(2023)]{Tcmap2023}
Tcmap (2023) Mamuchi Township Geographic Introduction. Available at: \url{http://www.tcmap.com.cn/shandong/yinanxian\_mamuchixiang.html} (accessed 20 April 2025).

\bibitem[Toutiao News(2022)]{Toutiao2022}
Toutiao News (2022) Rural revitalization in Yimeng: Yinan County Mamuchi Township—``Small crops'' contribute to ``great achievements'' in rural revitalization. Available at: \url{https://www.toutiao.com/article/7159506373804622344/?{\&}source=m\_redirect} (accessed 20 April 2025).

\bibitem[Webber SC, Porter MM and Menec VH(2010)]{WebberEtAl2010}
Webber SC, Porter MM and Menec VH (2010) Mobility in older adults: A comprehensive framework. The Gerontologist 50(4): 443–450.

% \bibitem[Wiles JL, Leibing A, Guberman N, Reeve J and Allen RE(2012)]{WilesEtAl2012}
% Wiles JL, Leibing A, Guberman N, Reeve J and Allen RE (2012) The meaning of ``aging in place'' to older people. The Gerontologist 52(3): 357–366.

\bibitem[World Health Organization(2007)]{WHO2007}
World Health Organization (2007) Global age-friendly cities: A guide. Geneva: WHO Press.

% \bibitem[World Health Organization(2017)]{WHO2017}
% World Health Organization (2017) Environmental health in emergencies: Solar ultraviolet radiation. Available at: \url{https://www.who.int/uv} (accessed 20 April 2025).

\bibitem[Yinan County Government(2018)]{Yinan2018}
Yinan County Government (2018) Survey report on regional public brands of agricultural products in Yinan County (Part 2). Available at: \url{http://www.yinan.gov.cn/info/1564/46850.htm} (accessed 20 April 2025).

\bibitem[Yinan County Government(2021)]{Yinan2021}
Yinan County Government (2021) Mamuchi Township Bama Village: Bits of beauty quietly bloom as picturesque rural scenery. Available at: \url{http://www.yinan.gov.cn/info/1536/106527.htm} (accessed 20 April 2025).

\bibitem[Yinan County Government(2022)]{Yinan2022}
Yinan County Government (2022) Mamuchi Township, Yinan County: ``Small crop'' sweet potatoes contribute to ``great achievements'' in rural revitalization. Available at: \url{http://www.yinan.gov.cn/info/1514/120706.htm} (accessed 20 April 2025).

\bibitem[Yinan County Government(2023)]{Yinan2023}
Yinan County Government (2023) Mamuchi Township October agricultural economic information. Available at: \url{http://www.yinan.gov.cn/info/1514/140601.htm} (accessed 20 April 2025).

\bibitem[Zeng Y and Wang Z(2014)]{ZengWang2014}
Zeng Y and Wang Z (2014) Dynamics of family and elderly living arrangements in China: New lessons learned from the 2000 census. China Review 14(2): 9–38.

\bibitem[Zhan HJ and Montgomery RJV(2003)]{ZhanMontgomery2003}
Zhan HJ and Montgomery RJV (2003) Gender and elder care in China: The influence of filial piety and structural constraints. Gender \& Society 17(2): 209–229.

\bibitem[Zhang X and Pan M(2020)]{ZhangPan2020}
Zhang X and Pan M (2020) Emerging rural spatial restructuring regimes in China: A tale of three transitional villages in the urban fringe. Journal of Rural Studies 93: 287-300.

\bibitem[Zhang Y, Chen M and Wang S(2023)]{ZhangEtAl2023a}
Zhang Y, Chen M and Wang S (2023) Different impact on health outcomes of long-term care insurance between urban and rural older residents in China. Scientific Reports 13: 27576.

\bibitem[Zhang Y, Guo Y and Xu H(2019)]{ZhangEtAl2019}
Zhang Y, Guo Y and Xu H (2019) Terrain and the rural elderly: Spatial accessibility and health disparity in mountainous regions of western China. Health \& Place 58: 102142.

\end{thebibliography}
\end{document}